\documentclass[doublecol]{epl2}

\title{Stopping in central Pb + Pb collisions at SPS energies and beyond}
\shorttitle{Stopping in central Pb+Pb at SPS energies} %Insert here a short version of the title if it exceeds 70 characters

\author{Yacine Mehtar-Tani\inst{1} \and Georg Wolschin\inst{2}}
\shortauthor{Y. Mehtar-Tani and G. Wolschin}

\institute{                    
  \inst{1} Departamento de F\'i˜sica de Part\'iculas and IGFAE, Universidade de Santiago de Compostela, Spain\\
  \inst{2} Institut f{\"ur} Theoretische 
Physik
der Universit{\"a}t Heidelberg, 
        Philosophenweg 16,  
        D-69120 Heidelberg, Germany
}

\abstract{We investigate stopping and baryon transport in central relativistic Pb + Pb and Au + Au collisions. 
At energies reached at the CERN Super Proton Synchrotron ($\sqrt {s_{NN}}$ = 6.3--17.3 GeV) and at RHIC (62.4 GeV), we determine the fragmentation-peak positions from the data. The resulting linear growth of the peak positions with beam rapidity is in agreement with our results from a QCD-based approach that accounts for gluon saturation. No discontinuities in the net-proton fragmentation peak positions occur in the expected transition region from partons to hadrons at  6--10 GeV.}

\begin{document}

\maketitle

\section{Introduction}

The search for signatures of the cross-over of the nuclear medium at high temperature and low chemical potential from a confined to a deconfined phase in relativistic heavy-ion collisions as predicted by lattice QCD \cite{aok09}, and for a first-oder phase transition at lower temperatures -- with a critical point at intermediate temperatures --
has a varied history. Some of the decisive observables in this context such as
strangeness enhancement \cite{koc86} have been discussed since more than 20 years.

Energy scans of such observables are indeed suitable to try to detect the threshold energy for the QCD phase transition. As an early example, a maximum  in the \(<K^{+}>/<\pi^{+}>\) ratio was found by the NA49 collaboration 
in central Pb + Pb collisions at \(\sqrt{s_{NN}} \simeq 8\) GeV which is ``... consistent with the hypothesis that a transient state of deconfined matter is created above these energies" \cite{afa02}. There exist, however, other interpretations of these data. In particular, the maximum is consistent with the decreasing net-baryon value combined with increasing $K\bar{K}$ pair production at rising \(\sqrt{s_{NN}}\)  \cite{adl04}.

The ongoing STAR beam energy scan program BES at RHIC \cite{ody10} tests the low-energy range
\(\sqrt{s_{NN}} = 5-50\) GeV in Au + Au collisions with the aim to look for specific signatures of a critical point in fluctuations and collective effects using data for identified particles from a single detector. It is also expected to see the disappearance of partonic effects at lower energies.

For \(\phi\)-Meson production, data from NA49 and STAR exist already for central
Cu + Cu, Au + Au and Pb + Pb collisions in the energy range $6.3-200$ GeV. 
It was argued in \cite{cha10} that these data ``indicate a threshold energy for the confinement-deconfinement phase transition" at \(\sqrt{s_{NN}} = 15.7 \pm 8.1\) GeV.

It has also been proposed to look for signatures of the deconfinement threshold in the net-baryon ($B-\bar{B}$) or net-proton rapidity distributions. As opposed to distributions of produced charged hadrons which are close to statistical equilibrium, these are clearly nonequilibrium distributions with two fragmentation peaks in rapidity space that move away from midrapidity with increasing c.m. energy \cite{bea04,wol99}. For a thorough test of theoretical models one needs the full rapidity distribution, not just the midrapidity region.

In \cite{wol03,wol06} it was shown on the basis of a Relativistic Diffusion Model \cite{wol99} that the net-proton rapidity distributions in central Au + Au at RHIC energies of 200 GeV can not be understood as being due to hadronic diffusion processes, and that the midrapidity region is populated through partonic processes that indicate deconfinement. In contrast, at AGS and low SPS energies the measured net-proton rapidity distribution can be modeled precisely within the hadronic diffusion description including a prediction of the correct energy dependence.

The energy dependence of the net-proton rapidity distributions at AGS and SPS energies has also been discussed by \cite{iva10} with respect to signs for a transition to the quark-gluon phase: A possible non-monotonous behaviour of the curvature of this distribution at midrapidity as function of the incident energy was investigated. There is an indication for such a behaviour in the region $\sqrt {s_{NN}}$ = 4--10 GeV from three-fluid dynamic calculations with an equation of state that involves a first-order transition \cite{iva10}, but it is not clear or even doubtful whether the alleged curvature anomaly is supported by the data with sufficient statistical significance.

In this Letter we investigate the net-proton distributions for central Pb + Pb collisions in the CERN Super Proton Synchrotron energy region $\sqrt {s_{NN}}$ = 6.3--17.3 GeV, and central Au + Au at RHIC energies of 62.4 GeV. 

We have made a detailed analysis of stopping in Pb + Pb collsions at 17.3 GeV with NA 49 data from \cite{app99,blu07}, and of Au + Au collisions at RHIC energies with BRAHMS data \cite{ars09} already in
\cite{mtw09,mtwc09}. Our partonic model accounts for interactions of the valence quarks with the gluon condensate in the respective other nucleus. It is not meant to fully account for stopping at considerably lower energies. The aim is here to look for indications of the hadron-parton transition in net-proton rapidity distributions at SPS energies through deviations between model results and data at low energies, or discontinuities as function of energy. 

With this aim in mind, we use the fragmentation-peak position as function of beam rapidity, or center of mass energy, as an observable. Since it can be determined quite accurately from the data at low energies, and is expected to depend linearly on the beam rapidity from the partonic model, it is likely to be a more precise indicator than the mean rapidity loss that had been discussed in \cite{vid95}. The mean rapidity loss deviates from a linear dependence on beam rapidity at RHIC energies and beyond, in accordance with our partonic model \cite{mtw09,mtwc09}.

We reconsider essential ingredients of the model in the next section, where the linear dependence of the fragmentation-peak positions on beam rapidity that the model predicts is emphasized. Then we determine the fragmentation-peak positions in analyses of the SPS and RHIC data, and find good agreement with the model predictions, but no evidence for discontinuities. We draw the conclusions in the last section.

\begin{figure}
\includegraphics[width=7.85cm]{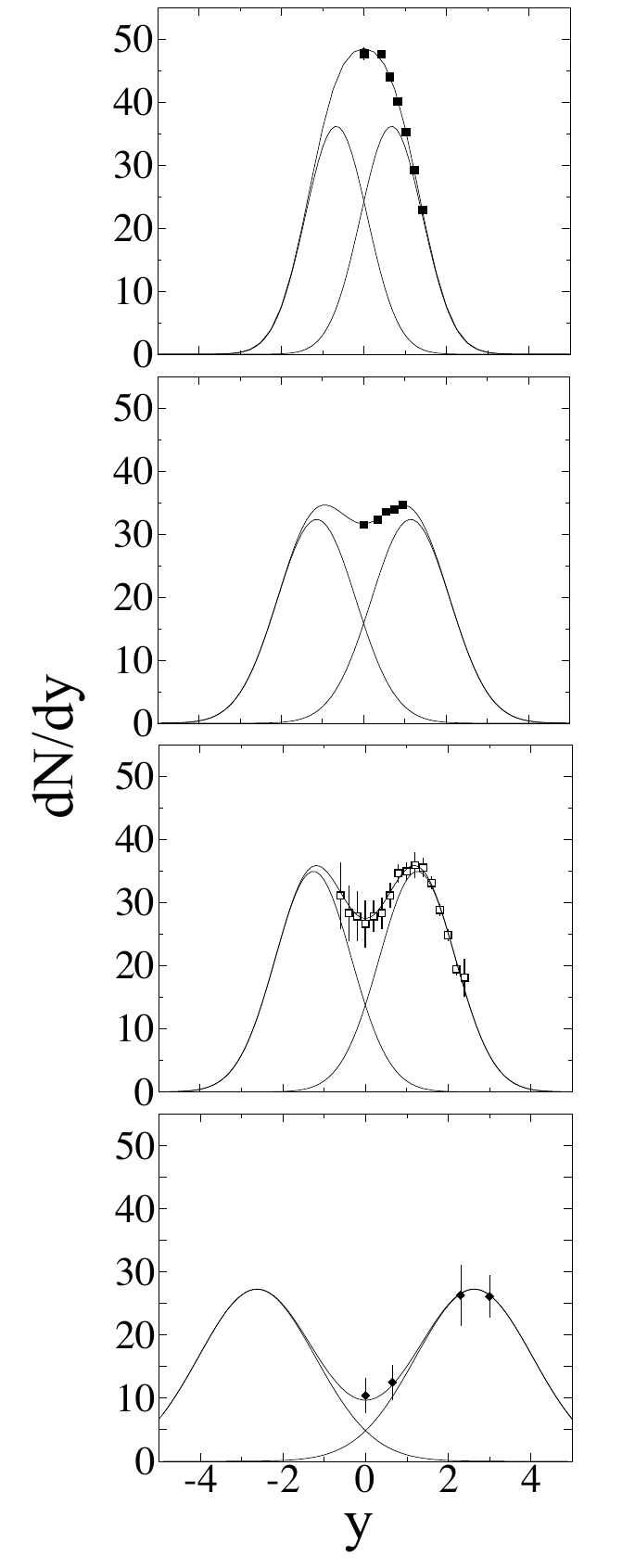}
%\onefigure{epl-template.eps}
\caption{Double-gaussian fits of the net-proton rapidity distributions in Pb + Pb  (from top to bottom) at $\sqrt{s_{NN}}$ = 6.3, 12.3 and 17.3 GeV, and in Au + Au at 62.4 GeV
%6.3, 7.6, 8.8, 12.3 and 17.3 GeV 
allow to determine the positions of the fragmentation peaks in rapidity space.
Preliminary Pb data points are from the NA49 collaboration \cite{blu07,blu08} using a conversion factor 82/208 from baryons to protons. At 17.3 GeV the data shown here
% (open symbols) 
are from the previous NA49-analysis \cite{app99}. Au data in the bottom panel are from BRAHMS \cite{ars09}. Error bars are smaller than the point sizes unless shown. See  table~\ref{tab.1} for results at all investigated energies, including the new Pb data at 17.3 GeV.}
\label{fig1}
\end{figure}

%\begin{figure}
%\includegraphics[width=8.8cm]{fig2.pdf}
%\onefigure{epl-template.eps}
%\caption{(Preliminary)Net-proton rapidity distributions in central Pb + Pb at $\sqrt{s_{NN}}$ = 6.3, 12.3 and %17.3 GeV from  QCD-calculations in the model of \cite{mtw09,mtwc09} extended here to lower energies
%are compared with NA49 data \cite{app99,blu07}. The model calculations produce slightly larger %fragmentation-peak positions at low energies, corresponding values are given in table~\ref{tab.1}. They %show no discontinuity in the peak positions as functions of energy.}
%\label{fig2}
%\end{figure}

\begin{figure}
\includegraphics[width=8.7cm]{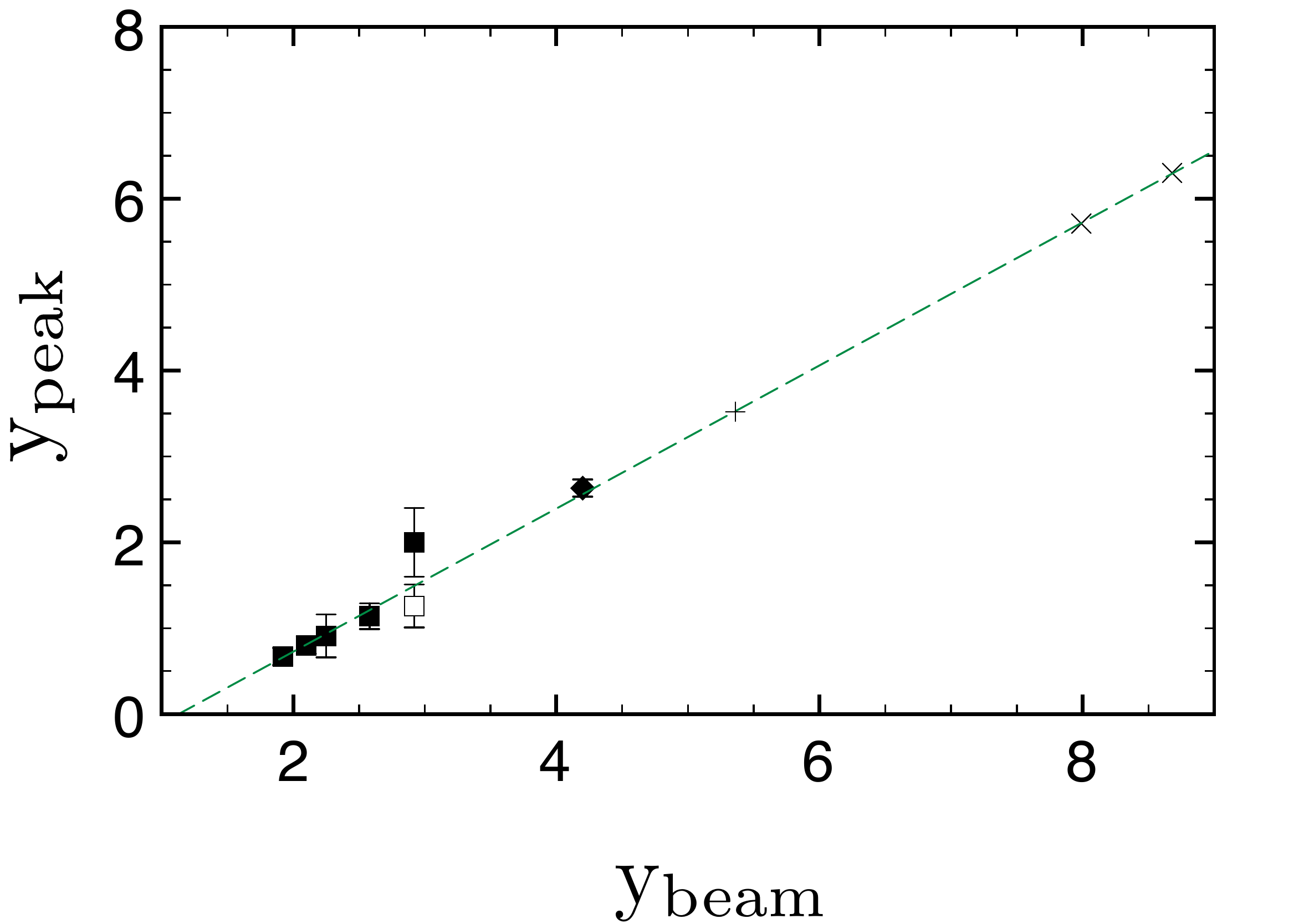}
%\onefigure{epl-template.eps}
\caption{Peak positions of the net-proton rapidity distributions in Pb + Pb (black squares) and Au + Au (black diamond) as function of the beam rapidity, determined from double-gaussian fits of the NA49 \cite{blu07,blu08} and RHIC data \cite {ars09}, see table~\ref{tab.1}.
% for numerical results.
 The open square is based on older NA49 data at 17.3 GeV \cite{app99}.
%The QCD-based calculation (solid)  deviate slightly (dashed line) from the data at %$y_b\approx$ 2.5--2.9,
%or $\sqrt{s}\approx$ 12--17 GeV. 
%Our QCD-based calculation 
%(parameters adjusted to 200 GeV Au + Au at RHIC, upper black diamond) has a 
The slope agrees well with the analytical expression 
eq.(\ref{eq:peak}) for $\lambda=0.2$,
% and const=$-0.2$,
dashed line. The cross refers to the calculated peak position of  Au + Au at 200 GeV, the inclined crosses to Pb + Pb at LHC energies of 2.76 and 5.52 TeV.}
\label{fig3}
\end{figure}

\begin{table}
\caption{Energies and beam rapidities for Pb + Pb (Au + Au$^a$), and corresponding positions and widths of the net-proton fragmentation peaks from double-gaussian fits. 
%Low-energy results 
The fit results $y^{fit}_{peak}$ and $\Gamma_{peak}^{fit}$ (FWHM) with estimated error bars 
%up to 62.4 GeV are from double-gaussian fits of 
refer to the NA49 data \cite{blu07,blu08} and to the BRAHMS data \cite{ars09}. Results marked as $^b$ are from a fit to previous NA49 data \cite{app99}.}
%The peak positions
%$y^{theor}_{peak}$  are calculated numerically in our theoretical partonic model at RHIC energies of 62.4 and %200 GeV. 
%are in good agreement with the BRAHMS data \cite{bea04,ars09}. 
%Also shown are Pb + Pb predictions at LHC energies of 0.35, 2.76 and 5.52 TeV(*).} 
%The last column shows the difference between the experimental peak positions, and the results %in the %partonic model. The errors bars for the low-energy data are smaller than the size of the %points.} 
\label{tab.1}
\begin{center}
\begin{tabular}{lllcr}
\hline\\
$\sqrt{s_{NN}} $&$|y_{beam}|$ & $|y^{fit}_{peak}|$&$\Gamma_{peak}^{fit}$\\
     (GeV)\\\\
\hline\\
      6.3 &1.92& 0.67$\pm 0.10$&1.78$\pm 0.2$\\
      7.6 & 2.10& 0.80$\pm 0.10$&2.08$\pm 0.2$\\
      8.8 & 2.25& 0.91$\pm 0.25$&2.19$\pm 0.5$\\
      12.3 & 2.58& 1.14$\pm 0.15$&2.25$\pm 0.3$\\
      17.3 & 2.92& 2.00$\pm 0.40$&3.11$\pm 0.6$\\
      17.3$^b$& 2.92& 1.26$\pm 0.25$&2.17$\pm 0.4$\\
   62.4$^a$& 4.20& 2.63$\pm 0.10$&3.33$\pm 0.2$\\\\
 %  200$^a$& 5.36& 3.51\\\\
%354.8*& 5.93& 4.0\\
% 2759.6*& 7.99& 5.7\\
%  5519.2*& 8.68& 6.3\\
\hline
\end{tabular}
\end{center}
\end{table}

\begin{figure}
\includegraphics[width=8.7cm]{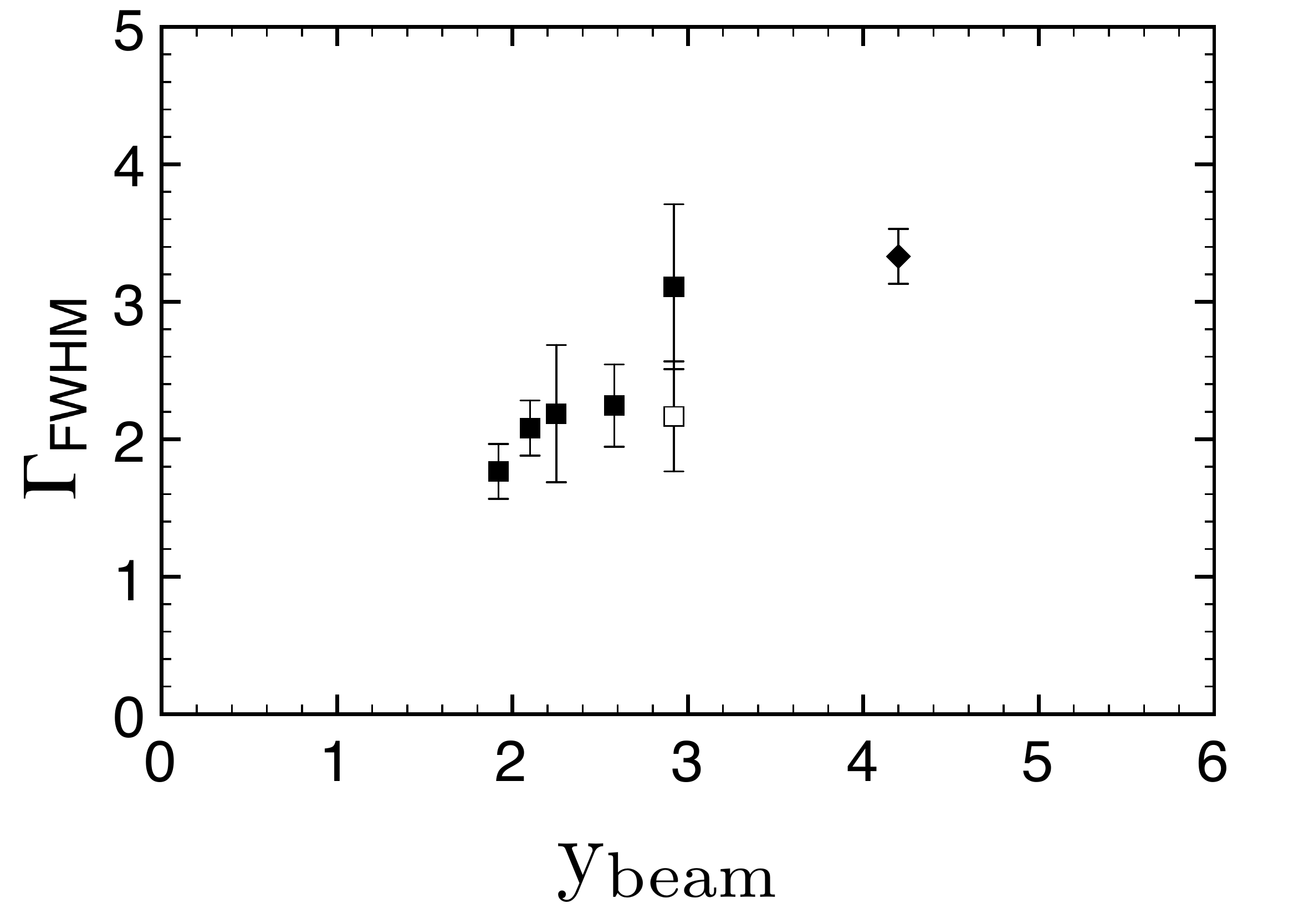}
%\onefigure{epl-template.eps}
%\vspace{.1%cm}
\caption{Widths (FWHM) of the net-proton fragmentation-peak rapidity distributions in Pb + Pb (squares) and Au + Au (diamond) as functions of the beam rapidity based on double-gaussian fits (fig.~\ref{fig1}) of the available SPS data (black squares) from NA49 \cite{app99,blu07,blu08}, and of RHIC data from BRAHMS \cite {bea04,ars09}. The widths increase at SPS energies. At high RHIC energies and at LHC, a constant value is expected from our model.}
\label{fig6}
\end{figure}

\section{Stopping in QCD}

The differential cross section for valence quark production in a high-energy nucleus-nucleus collision
is calculated from \cite{kha04,bai06,dum06}
\begin{equation} \label{eq:crossGS}
\frac{dN}{d^2p_Tdy}= \frac{1}{(2\pi)^2 } \frac{1}{ p_T^2}\;x_1q_v(x_{1},Q_{f})\;\varphi\left(x_2,p_T\right),
\end{equation}
where \(p_T\) is the transverse momentum of the produced quark, and \(y\) its rapidity. The longitudinal momentum fractions carried, respectively, by the valence quark in the projectile and the soft gluon in the target are \(x_1=p_T/\sqrt{s}\exp(y)\) and \(x_2=p_T/\sqrt{s}\exp(-y)\). The factorization scale is set equal to the transverse momentum, \(Q_{f}\equiv p_T\). The valence-quark distribution is $q_v(x_{1},Q_{f})$. 
We have discussed the gluon distribution \(\varphi(x,p_T)\) and details of the overall model in
\cite{mtw09,mtwc09}.

One important prediction of the color glass condensate theory \cite{gri83,mue86,bla87,mcl94} is 
geometric scaling: the gluon distribution depends on $x$ and $p_T$ 
only through the scaling variable $p_T^2/Q_s^2(x)$, where $Q_s^2(x)=A^{1/3} Q_0^2\;x^{-\lambda}$, $A$ is the mass number, $Q_0$ sets the dimension, and $\lambda$ is the saturation-scale exponent. This has been confirmed experimentally for $A$ = 1 in deep inelastic $e + p$ scattering 
at HERA \cite{sta01}. There the fit value $\lambda = 0.2-0.3$ agrees with theoretical estimates based on  the Balitsky-Kovchegov equation \cite{bal96,kov99} that include a running coupling \cite{bal07,kow07,alb07}.

To show that the net-baryon distribution in relativistic heavy-ion collisions also reflects the geometric scaling of the gluon distribution, the following change of variables
has been proposed in \cite{mtw09}: 
%\begin{equation}
\(x\equiv x_1,\;\;x_2\equiv x\;e^{-2y},\;\; p_T^2\equiv x^2 s\; e^{-2y}.\)
%\end{equation}
Thus after integrating over $p_T$, the rapidity distribution can be rewritten as 
\begin{equation}
\frac{dN}{dy}(\tau)=\frac{C}{2\pi}\int_0^1\frac{dx}{x}\;xq_v(x) \;\varphi(x^{2+\lambda} 
e^\tau),
\label{eq:GSyield}
\end{equation}
where $\tau=\ln (s/Q_0^2) - \ln A^{1/3} - 2(1+\lambda)\,y$ is the corresponding scaling variable and $C$ is the overall normalization that contains the fragmentation function \cite{mtwc09}, and is fitted here to the experimental yields. Hence, the 
net-baryon rapidity density in the peak region is a function of a single 
scaling variable $\tau$, which relates the energy dependence to the rapidity and
mass number dependence. 
%In the fragmentation region,
%the valence quark distribution is only very weakly dependent on $Q_{f}$.

For the fragmentation-peak position, we have derived in \cite{mtw09} the analytical result
\begin{equation}\label{eq:peak}
y_{peak}=\frac{1}{1+\lambda}\left(y_{beam}-\ln 
A^{1/6}\right)+const.
\end{equation}

It is expected to be applicable at RHIC energies and beyond, but here we test the validity of the predicted linear growth with $y_{beam}$ at SPS energies. At very low energies deviations will occur due to kinematical constraints in the limit $y_{beam}\rightarrow 0$ where $y_{peak}\rightarrow 0$.

\section{Results}

Using the model as outlined in the previous section, we have investigated baryon transport in central Pb + Pb collisions at SPS energies of $\sqrt {s_{NN}}$ = 6.3--17.3 GeV, and in central Au + Au at RHIC energies of 62.4 and 200 GeV. In the high-energy region, our parton-based approach gives good results for net-proton rapidity distributions \cite{mtw09,mtwc09} with respect to the available BRAHMS data \cite{bea04,ars09}. 

With the parameters fixed at 200 GeV ($\lambda = 0.2$, and $Q_0^2$ = 0.04 GeV$^2$), we have also used it for predictions at LHC energies of 2.76 and 5.52 TeV \cite{mtw10}. These theoretical results at LHC energies will, however, not be tested experimentally in the next years at or near the fragmentation peaks since particle identification at these energies is restricted so far to $|y|<2$.

In the region of  SPS and low RHIC energies, we have determined the net-proton fragmentation-peak positions $y_{peak}^{fit}$ from the preliminary data of the NA49 collaboration \cite{blu07,blu08} and of BRAHMS \cite{ars09} in double-gaussian fits, as shown in fig.~\ref{fig1} and table~\ref{tab.1}.
There is an uncertainty regarding the peak position at 17.3 GeV since the experiments from two different runs at this energy have produced different results \cite{app99,blu08}, with the 
more recent preliminary data not extending across the fragmentation-peak position. 

Different from the Au + Au data at 62.4 GeV, the experimental BRAHMS results at 200 GeV \cite{bea04} can not be reproduced in a double-gaussian fit to determine the fragmentation-peak positions. Instead, an additional midrapidity source for net protons appears that had been interpreted in \cite{wol03} as being due to partonic contributions to the net-proton yield. It is related to valence quark -- gluon collsions since  gluon-gluon collisions do not contribute to net-baryon distributions.  Indeed the data are in good agreement with our partonic calculation \cite{mtw09,mtwc09} which had been calibrated at this energy.

%The net-proton fragmentation-peak position $y_{peak}^{theor}$ as calculated numerically at %these energies in our microscopic model without any further adaption of the parameters is %slightly larger than the low-energy NA49 data, %fig.~\ref{fig2} and 
%table~\ref{tab.1}:
%The scattering of valence quarks off the gluon condensate that is the basis of these calculations %favors slightly larger fragmentation-peak positions (smaller scattering angles) than the mostly %hadronic processes that actually take place in this energy domain. 
%The approximate analytical expression for the peak position as derived in \cite{mtw09} that is %shown as %dashed line in fig.~\ref{fig3} agrees well with the data.  

The peak positions determined from the fits as functions of beam rapidity are shown in fig.~\ref{fig3}.
% as functions of center-of-mass energy in fig.~\ref{fig5}.  
Here the beam rapidity is obtained from the center-of-mass energy as $y_{beam} = \ln(\sqrt{s_{NN}}/m_p)$ with the proton mass $m_p$. The peak positions are found to
increase linearly with beam rapidity at SPS and RHIC energies. The results of the double-gaussian fits are in surprisingly good agreement with the prediction of our microscopic model in eq.~(\ref{eq:peak}) 
for a saturation-scale exponent $\lambda=0.2$, 
an empirical constant $= -0.2$, and A=208 ($208^{1/6} \simeq 197^{1/6}$).
%and a constant $= - 0.2$, see dashed line in fig.~\ref{fig3}.
%Here the constant is empirical, it is a non-perturbative quantity and cannot be calculated analytically. The %agreement of its absolute value with $\lambda$ is probably accidential.
%The numerical calculations shown in table~\ref{tab.1}
%as shown in fig.~\ref{fig2}
%deviate slightly from this analytical result at low SPS energies where contributions from both %beams overlap, leading to small shifts of the peak positions to larger values of rapidity which %we list in table~\ref{tab.1} as $y_{peak}^{theor}$.

At SPS and RHIC energies, a saturation-scale exponent $\lambda=0.2$ clearly gives the best result for the
slope in comparison with the NA49 and BRAHMS data, in agreement with results of \cite{hir04,alb07}. 
At high RHIC energies of 200 GeV and LHC energies of 2.76 and 5.52 TeV which are also indicated in fig.~\ref{fig3}, a larger value of $\lambda$ may turn out to be more adequate, depending on the growth of the saturation scale with energy. 

A saturation-scale exponent $\lambda=0.2$ corresponds to a gluon saturation momentum of $Q_s \simeq 0.77$ GeV at $x$=0.01. When comparing this value to investigations of charged-hadron production such as in \cite{alb07}, which involve the gluon distribution in the adjoint representation of SU(3), one has to consider a rescaling of our net-baryon $Q_s^2$ by the color factor $N_C/C_F$ with $C_F = (N_C^2-1)/2N_C$ and $N_C = 3$, corresponding to a factor 9/4
\cite{mtw10}. This leads to $Q_s \simeq 1.2$ GeV at $x$=0.01.

There is no discontinuity in the peak positions as functions of center of mass energy, nor any disagreement between calculation and data analysis that might indicate the sudden onset of purely hadronic processes at lower energies.  
Hence, the transition from partonic to hadronic processes in the observable net-proton fragmentation-peak positions is gradual, without immediate signatures of a cross-over as predicted from lattice gauge theories for equilibrium systems of partons. 

The widths of the net-proton distribution functions as determined from the SPS and RHIC data
in the double-gaussian fits are shown in table~\ref{tab.1} and fig.~\ref{fig6}. Plotted are the full widths $\Gamma$ at half maximum (FWHM) of the individual fragmentation peaks, which are related to the standard deviations $\sigma$ through $\Gamma = \sqrt{8\ln2}\cdot\sigma$. Similar to the mean values, they increase with beam rapidity from SPS to RHIC energies. The widths will tend to the width of the Fermi distribution in rapidity space as $y_{beam}\rightarrow 0$.

An increase of the width with center-of-mass energy is  expected in the Relativistic Diffusion Model \cite{wol07} that has been used to interpret SPS and 62 GeV RHIC data. In the high-energy limit, our microscopic model predicts the width to stay constant, but again this result will be difficult to test at LHC energies.
The transition from an increasing width at SPS energies, to a constant width at higher RHIC energies and beyond, would be indicative of the transition from predominantly hadronic to partonic system behaviour.

%This result is in contrast to the claim in \cite{iva10} of cross-over signatures in net-proton %rapidity distributions in the energy region 4--10 GeV.

\section{Conclusion}

An analysis of net-proton rapidity  distributions in central Pb + Pb collisions at SPS c.m. energies 
of 6--17 GeV, and Au + Au at low RHIC energies of 62.4 GeV has been presented. The fragmentation-peak positions have been determined in double-gaussian fits of the available data. On the theoretical side, the investigation uses our microscopic approach based on quantum chromodynamics that analytically predicts the fragmentation-peak positions as function of center-of-mass energy, or beam rapidity ~\cite{mtw09,mtwc09}.

When tested at low energies, good agreement with the fragmentation-peak positions
determined from the data in the measured net-proton distributions occurs at c.m. energies of 6--62 GeV. The linear growth of the fragmentation-peak position with beam rapidity that our microscopic model predicts is indeed found in the SPS and RHIC data. 

There is no discontinuity as function of energy, nor a severe disagreement with the data that might have indicated the sudden onset of hadronic processes at low energies as consequence of a parton-hadron transition. Hence, the highly non-equilibrium net-baryon peak positions in rapidity space do not seem to be well-suited as signatures for the parton-hadron cross-over, or first-order phase transition at higher values of the chemical potential.

There are indications for a cross-over signature in net-proton distributions near midrapidity, as has been suggested in \cite{wol03,iva10}. However, there are probably better chances to establish the existence of the hadron-parton transition in observables that are much closer to statistical equilibrium, such as rapidity distributions and cross-sections of produced strange mesons. The fact that such observables approach statistical equilibrium much closer than net-baryon distributions makes it more plausible to compare with lattice results regarding the hadron-parton cross-over.

%See fig.~\ref{fig1}, table~\ref{tab.1} and eq.~(\ref{eq.1}).
%See also~\cite{afa02,iva10,cha10,mtw09,mtwc09,mtw10}.
%\begin{equation}
%\label{eq.1}
%0\neq1
%\end{equation}

\acknowledgments
This work is supported by the ExtreMe Matter Institute, EMMI.

\bibliographystyle{eplbib}
\bibliography{gw_epl}

\begin{thebibliography}{10}
\expandafter\ifx\csname url\endcsname\relax\def\url#1{\texttt{#1}}\fi

\bibitem{aok09}
\Name{Aoki Y. \etal} \REVIEW{Nucl. Phys. A}{830}{2009}{805c}.

\bibitem{koc86}
\Name{Koch P., Mueller B. \and Rafelski J.} \REVIEW{Phys. Rep.
  }{142}{1986}{167}.

\bibitem{afa02}
\Name{Afanasiev S.} \REVIEW{Phys. Rev. C}{66}{2002}{054902}.

\bibitem{adl04}
\Name{Adler C. \etal} \REVIEW{Phys. Lett. B}{595}{2004}{143}.

\bibitem{ody10}
\Name{Odyniec G.} \REVIEW{J. Phys. G: Nucl. Part. Phys. }{37}{2010}{094028}.

\bibitem{cha10}
\Name{Chaudhuri A.~K.} \REVIEW{Phys. Lett. B}{690}{2010}{261}.

\bibitem{bea04}
\Name{Bearden I.~G. \etal} \REVIEW{Phys. Rev. Lett. }{93}{2004}{102301}.

\bibitem{wol99}
\Name{Wolschin G.} \REVIEW{Eur. Phys. J. A }{5}{1999}{85}.

\bibitem{wol03}
\Name{Wolschin G.} \REVIEW{Phys. Lett. B}{569}{2003}{67}.

\bibitem{wol06}
\Name{Wolschin G.} \REVIEW{Europhys. Lett. }{74}{2006}{29}.

\bibitem{iva10}
\Name{Ivanov Y.~B.} \REVIEW{Phys. Lett. B}{690}{2010}{358}.

\bibitem{app99}
\Name{Appelsh{\"a}user H. \etal} \REVIEW{Phys. Rev. Lett. }{82}{1999}{2471}.

\bibitem{blu07}
\Name{Blume C. \etal} \REVIEW{J. Phys. G }{34}{2007}{951}.

\bibitem{ars09}
\Name{Arsene I.~C. \etal} \REVIEW{Phys. Lett. B}{677}{2009}{267}.

\bibitem{mtw09}
\Name{Mehtar-Tani Y. \and Wolschin G.} \REVIEW{Phys. Rev. Lett.
  }{102}{2009}{182301}.

\bibitem{mtwc09}
\Name{Mehtar-Tani Y. \and Wolschin G.} \REVIEW{Phys. Rev. C
  }{80}{2009}{054905}.

\bibitem{vid95}
\Name{Videbaek F. \and Hansen O.} \REVIEW{Phys. Rev. C }{52}{1995}{2684}.

\bibitem{blu08}
\Name{Blume C. \etal} \REVIEW{PoS (Confinement) }{8}{2008}{110}.

\bibitem{kha04}
\Name{Kharzeev D., Kovchegov Y.~V. \and Tuchin K.} \REVIEW{Phys. Lett.
  B}{599}{2004}{23}.

\bibitem{bai06}
\Name{Baier R., Mehtar-Tani Y. \and Schiff D.} \REVIEW{Nucl. Phys.
  A}{764}{2006}{515}.

\bibitem{dum06}
\Name{Dumitru A., Hayashigaki A. \and Jalilian-Marian J.} \REVIEW{Nucl. Phys.
  A}{765}{2006}{464}.

\bibitem{gri83}
\Name{Gribov L.~V., Levin E.~M. \and Ryskin M.~G.} \REVIEW{Phys. Rep.
  }{100}{1983}{1}.

\bibitem{mue86}
\Name{Mueller A.~H. \and Qiu J.} \REVIEW{Nucl. Phys. B}{268}{1986}{427}.

\bibitem{bla87}
\Name{Blaizot J.~P. \and Mueller A.~H.} \REVIEW{Nucl. Phys. B}{289}{1987}{847}.

\bibitem{mcl94}
\Name{McLerran L. \and Venugopalan R.} \REVIEW{Phys. Rev. D }{49}{1994}{2233}.

\bibitem{sta01}
\Name{Sta\'sto A.~M., Golec-Biernat K. \and Kwieci\'nski J.} \REVIEW{Phys. Rev.
  Lett. }{86}{2001}{596}.

\bibitem{bal96}
\Name{Balitsky I.} \REVIEW{Nucl. Phys. B}{463}{1996}{99}.

\bibitem{kov99}
\Name{Kovchegov Y.~V.} \REVIEW{Phys. Rev. D }{60}{1999}{034008}.

\bibitem{bal07}
\Name{Balitsky I.} \REVIEW{Phys. Rev. D }{75}{2007}{014001}.

\bibitem{kow07}
\Name{Kovchegov Y.~V. \and Weigert H.} \REVIEW{Nucl. Phys. A}{784}{2007}{188}.

\bibitem{alb07}
\Name{Albacete J.~L.} \REVIEW{Phys. Rev. Lett. }{99}{2007}{262301}.

\bibitem{mtw10}
\Name{Mehtar-Tani Y. \and Wolschin G.} \REVIEW{Phys. Lett. B }{688}{2010}{174}.

\bibitem{hir04}
\Name{Hirano T. \and Nara Y.} \REVIEW{Nucl. Phys. A }{743}{2004}{305}.

\bibitem{wol07}
\Name{Wolschin G.} \REVIEW{Prog. Part. Nucl. Phys. }{59}{2007}{374}.

\end{thebibliography}

\end{document}